\documentclass[12pt]{JHEP3} 
\usepackage{caption2}
\usepackage{cite}
\usepackage[dvips]{graphicx}
\usepackage[latin1]{inputenc}
\usepackage[squaren]{SIunits}
\usepackage{amsmath, amstext, amsfonts, mathrsfs}

\DeclareMathOperator{\tr}{tr}

\makeatletter
\newcommand{\fmslash}[2][0mu]{%
  \mathchoice
    {\fmsl@sh\displaystyle{#1}{#2}}%
    {\fmsl@sh\textstyle{#1}{#2}}%
    {\fmsl@sh\scriptstyle{#1}{#2}}%
    {\fmsl@sh\scriptscriptstyle{#1}{#2}}}
\newcommand{\fmsl@sh}[3]{%
  \m@th\ooalign{$\hfil#1\mkern#2/\hfil$\crcr$#1#3$}}
\title{UV/IR mixing in noncommutative QED defined by Seiberg-Witten map}

\author{Peter Schupp, Jiangyang You\\
School of Engineering and Science\\
Jacobs University\\
Campusring 1\\
28759 Bremen, Germany\\
{\tt E-mail: p.schupp@jacobs-university.de,
j.you@jacobs-university.de}} \preprint{hep-th/yymmnnn} \abstract{
Noncommutative gauge theories defined via Seiberg-Witten map have
desirable properties that theories defined directly in terms of
noncommutative fields lack, covariance and unrestricted choice of
gauge group and charge being among them, but non\-perturbative
results in the deformation parameter $\theta$ are hard to obtain. In
this article we use a $\theta$-exact approach to study UV/IR mixing
in a noncommutative quantum electrodynamics (NCQED) model defined
via Seiberg-Witten map. The fermion contribution of the one loop
correction to the photon propagator is computed and it is found that
it gives the same UV/IR mixing term as a NCQED model without
Seiberg-Witten map.}
\begin{document}
\section{Introduction}
Noncommutative quantum electrodynamics (NCQED) is usually defined in
analogy to Yang-Mills theory with matrix multiplication replaced by
star products. The resulting action is invariant under
noncommutative gauge transformations. Such models appear quite
naturally in certain limits of string theory in the presence of a
background $B$-field~\cite{Seiberg:1999vs}, they can also be used to
gain some understanding about phenomenological implications of a
quantum structure of space\-time. One of the particularly intriguing
effects is UV/IR-mixing, an interrelation between short and
long-distance scales that is absent in ordinary quantum field
theory. There are however some problems with this simple definition
of NCQED: (1) The possible choices of charges for particles are
restricted to $\pm 1$ or $0$ times a fixed unit of charge and in the
non\-abelian case the choice of structure group is limited to $U(N)$
in the fundamental representation. (2) An ordinary gauge field
$a_\mu(x)$ transforms like a vector under a change of coordinates,
$a'_\mu(x') = \partial x^\nu / \partial x'^\mu a_\nu(x)$, while for
the fields $A_\mu(x)$ of NCQED this holds only for rigid, affine
coordinate changes~\cite{Jurco:2001my,Jackiw:2001jb}. The solution
to both problems is an alternative approach to noncommutative gauge
theory based on Seiberg-Witten maps. This approach to noncommutative
gauge theory (especially, the noncommutative extension of
non\-abelian gauge theories) has been established for quite some
time~\cite{Madore:2000en,Bichl:2001gu,Jurco:2001rq}. The idea is to
consider noncommutative gauge fields $A_\mu$ and gauge
transformation parameters $\Lambda$ that are valued in the
enveloping algebra of the gauge group and can be expressed in terms
of the ordinary gauge field $a_\mu$, gauge parameter $\lambda$ and
the noncommutative parameter $\theta^{\mu\nu}$ in such a way that an
ordinary gauge transformations of $a_\mu$ induces a non-commutative
gauge transformation of $A_\mu[a]$ with non-commutative gauge
parameter $\Lambda[\lambda,a]$.

Comparing to the simpler formulation in which the gauge group is
directly deformed by replacing the normal product of group elements
with the Moyal-Weyl star product, the Seiberg-Witten map approach
removes the restrictions on the gauge group and charge and allows
the construction of more realistic models. The noncommutative action
can be treated as a complicated action written in terms of ordinary
fields, which when expanded in the powers of the non\-commutativity
parameter $\theta$ gives the usual commutative action (both free and
interacting parts) at zeroth order in $\theta$ plus higher order
non-commutative corrections. Therefore, such a theory can be
considered to be a minimal noncommutative extension of the
corresponding commutative model. Following this line, a minimal
noncommutative extension of the standard model has been
established\cite{Calmet:2001na,Melic:2005fm,Melic:2005am} and
influences to particle physics have been studied up to loop level in
low orders of $\theta$ \cite{Buric:2007qx,Alboteanu:2007zz}. Besides
being useful for phenomenology, the $\theta$-expansion was also
shown to be improving the renormalizability of the noncommutative
gauge
theory\cite{Bichl:2001cq,Buric:2002gm,Buric:2005xe,Buric:2006wm}.
The photon self energy is renormalizable up to any finite order of
$\theta$\cite{Bichl:2001cq} (with the sacrifice of introducing an
infinite number of coupling constant from the freedom/ambiguity
within the Seiberg-Witten map).

Although the $\theta$-expansion method works nicely in model
building, crucial non\-perturbative information is lost due to the
cut off at finite order of $\theta$. It is long known that in the
noncommutative field
theories\cite{Minwalla:1999px,Hayakawa:1999yt,Hayakawa:1999zf,Matusis:2000jf},
the Moyal-Weyl star product results in a nontrivial phase factor for
the Fourier modes when two functions are multiplied together. Such a
phase, when it appears in loop calculation, regulates the
ultraviolet divergence in the one loop two point function of both
noncommutative $\phi^4$ and noncommutative quantum electrodynamics
(NCQED) but introduces an infrared divergent term of the form
$1/(p\theta\theta\!p)$. As the nontrivial phase factor appears only
when all orders of $\theta$ in the star product are summed over,
this effect does not show up in the noncommutative gauge theories
defined by the Seiberg-Witten map approach when it is studied using
the $\theta$-expansion method, (thus it is sometimes claimed that
such a theory is free of UV/IR mixing). However, as already
suggested in some very early
papers\cite{Mehen:2000vs,Jurco:2001my,Jurco:2001rq}, the
$\theta$-expansion is not the only possible way of expressing the
Seiberg-Witten map. As the noncommutative gauge field $A_\mu$ is a
function of both the ordinary field $a_\mu$ and the
non\-commutativity parameter $\theta^{ij}$, one can, instead of
expanding $A_\mu$ in power of $\theta$, expand it in powers of
$a_\mu$. The first several orders of the expansion can be written in
a simple form by introducing certain generalized star
products\cite{Jurco:2001rq,Mehen:2000vs}. Such an expansion enables
us to treat all orders of $\theta$ at once in each interaction
vertex, thereby allowing us to compute non\-perturbative results. In
this article we are going to use this expansion to compute the
fermion one loop correction to the photon two point function of a
NCQED model defined by Seiberg-Witten map. We will see that UV/IR
mixing will still arise via the nontrivial phase factors, hence the
absence of UV/IR mixing in the Seiberg-Witten map approach to
noncommutative gauge theory so far has been really a technical
artifact of the perturbative $\theta$-expansion method, but not a
feature of the theory itself.

\section{The model}
For simplicity we consider a NCQED model with a $U(1)$ gauge field
$A_\mu$ and a fermion field $\Psi$ which lives in the adjoint
representation of the noncommutative gauge group $U(1)_{\star}$. The
action is as following\footnote{We use a Minkowskian signature here.
In the next section we allow a Wick rotation, thus the result is
actually on noncommutative $R^4$. This procedure is the same as
taken in \cite{Hayakawa:1999yt,Hayakawa:1999zf,Matusis:2000jf}, but
differs from a procedure where the action is directly written down
in $R^4$.}
\begin{equation}
S=\int-\frac{1}{4}F^{\mu\nu}F_{\mu\nu}+i\bar\Psi
\fmslash{\mathcal{D}}\Psi
\end{equation}
with
\begin{gather*}
D_\mu\Psi=\partial_\mu\Psi-i[A_\mu\stackrel{\star}{,}\Psi]\quad\mbox{and}\quad
F_{\mu\nu}=\partial_\mu A_\nu-\partial_\nu
A_\mu-i[A_\mu\stackrel{\star}{,}A_\nu]
\end{gather*}

The $\theta$-exact Seiberg-Witten map can be obtained in several
ways: From the closed formula derived using deformation quantization
based on Kontsevich formality map\cite{Jurco:2001my}, by the
relationship between open Wilson lines in the commutative and
noncommutative picture\cite{Mehen:2000vs}, or by a direct recursive
computation using consistency conditions. The computation of the one
loop two-point function requires fully $\theta$-exact  interaction
vertices up to four external legs, i.e.\ one needs the
$\theta$-exact Seiberg-Witten map of $A_\mu$ up to third order in
$a_\mu$.  This has been computed in its inverse form ($a_\mu$ in
terms of $A_\mu$ up to $A^3$) in \cite{Mehen:2000vs}. Here we simply
take the  inverse of this result by matching the trivial identity
$a_\mu(A_\mu(a_\mu))=a_\mu$ order by order, resulting in
\begin{equation}
\begin{split}
A_\mu&=\,a_\mu-\frac{1}{2}\theta^{ij}a_i\star_2(\partial_j
a_\mu+f_{j\mu})+\frac{1}{2}\theta^{ij}\theta^{kl}\bigg\{\frac{1}{2}(a_k\star_2(\partial_l
a_i+f_{li}))\star_2(\partial_j
a_\mu+f_{j\mu})\\&+a_i\star_2(\partial_j (a_k\star_2(\partial_l
a_\mu+f_{l\mu}))-\frac{1}{2}\partial_\mu (a_k\star_2(\partial_l
a_j+f_{lj})))-a_i\star_2(\partial_k a_j\star_2\partial_l
a_\mu)+\\&[a_i\partial_k a_\mu(\partial_j
a_l+f_{jl})-\partial_k\partial_i a_\mu a_j a_l+2\partial_k
a_i\partial_\mu a_j a_l]_{\star_3}\bigg\}+\mathcal O(A^4)
\end{split}
\end{equation}
where $\star, \star_2, \star_3$ are Moyal-Weyl star product and two
generalized star products:
\begin{gather}
f(x)\star g(x)=e^{\frac i2 \theta^{\mu\nu}{\frac{\partial}{\partial
y^\mu}}{\frac{\partial}{\partial
z^\nu}}}f(y)g(z)\bigg|_{x=y=z}\\f(x)\star_2
g(x)=\frac{\sin\frac{\partial_1\wedge
\partial_2}{2}}{\frac{\partial_1\wedge
\partial_2}{2}}f(x_1)g(x_2)\bigg|_{x_1=x_2=x}\\
[f(x)g(x)h(x)]_{\star_3}=\big[\frac{\sin(\frac{\partial_2\wedge
\partial_3}{2})\sin(\frac{\partial_1\wedge(\partial_2+\partial_3)}{2})}{\frac{(\partial_1+\partial_2)\wedge \partial_3}{2}\frac{\partial_1\wedge(\partial_2+\partial_3)}{2}}
+\{1\leftrightarrow 2\}\,\big]f(x_1)g(x_2)h(x_3)\bigg|_{x_i=x}
\end{gather}
where
\begin{gather}
\partial_1\wedge \partial_2=\theta^{ij}\frac{\partial}{\partial x_1^i}\frac{\partial}{\partial x_2^j}
\end{gather}

The expansion for a matter particle in the adjoint representation of
$U(1)_{\star}$ can be easily obtained by taking the linear part
(linear operator acting on $a_\mu$) in the expansion of $A_\mu$,
which leads to following result:
\begin{equation}
\begin{split}
\Psi&=\psi-\theta^{ij}
a_i\star_2\partial_j\psi+\frac{1}{2}\theta^{ij}\theta^{kl}\bigg\{(a_k\star_2(\partial_l
a_i+f_{li}))\star_2\partial_j\psi+2a_i\star_2(\partial_j(a_k\star_2\partial_l\psi))\\&-
a_i\star_2(\partial_k
a_j\star_2\partial_l\psi)-\big[a_i\partial_k\psi(\partial_j
a_l+f_{jl})-\partial_k\partial_i\psi a_j
a_l\big]_{\star_3}\bigg\}+\mathcal O(a^3)\psi
\end{split}
\end{equation}

Now the action can be expanded as following:
\begin{eqnarray}\label{ncqed:sw}
S=\int-\frac{1}{4}f^{\mu\nu}f_{\mu\nu}+i\bar\psi
\fmslash\partial\psi+L_{pp}+L_{pf}
\end{eqnarray}
$L_{pf}$ and $L_{pp}$ are photon-fermion and photon self interaction
terms, in this article we concentrate on the photon-fermion part, so
we write out $L_{pf}$ explicitly:
\begin{equation}\label{pf:int}
\begin{split}
L_{pf}&=\bar\psi\gamma^\mu[a_\mu\stackrel{\star}{,}\psi]+i(\theta^{ij}\partial_i\bar\psi
\star_2 a_j)\fmslash\partial\psi-i\bar\psi\star
\fmslash\partial(\theta^{ij}
a_i\star_2\partial_j\psi)+(\theta^{ij}\partial_i\bar\psi \star_2
a_j)\gamma^\mu[a_\mu\stackrel{\star}{,}\psi]\\&-\!\bar\psi\gamma^\mu[a_\mu\!\stackrel{\star}{,}\!\theta^{ij}
a_i\!\star_2\!\partial_j\psi]\!-\!\bar\psi\gamma^\mu[\frac{1}{2}\theta^{ij}a_i\!\star_2\!(\partial_j
a_\mu\!+\!f_{j\mu})\!\stackrel{\star}{,}\!\psi]\!-\!i(\theta^{ij}\partial_i\bar\psi
\!\star_2\!a_j)\fmslash\partial(\theta^{kl}
a_k\!\star_2\!\partial_l\psi)\\&+\frac{i}{2}\theta^{ij}\theta^{kl}\big((a_k\star_2(\partial_l
a_i+f_{li}))\star_2\partial_j\bar\psi+2a_i\star_2(\partial_j(a_k\star_2\partial_l\bar\psi))-
a_i\star_2(\partial_k
a_j\star_2\partial_l\bar\psi)\\&+\big[a_i\partial_k\bar\psi(\partial_j
a_l+f_{jl})-\partial_k\partial_i\bar\psi a_j a_l\big]_{\star_3}\big)
\fmslash\partial\psi+\frac{i}{2}\theta^{ij}\theta^{kl}\bar\psi\fmslash\partial\big((a_k\star_2(\partial_l
a_i+f_{li}))\star_2\partial_j\psi\\&+\!2a_i\!\star_2\!(\partial_j(a_k\!\star_2\!\partial_l\psi))\!-\!a_i\!\star_2\!(\partial_k
a_j\!\star_2\!\partial_l\psi)\!+\!\frac{1}{2}\theta^{ij}\theta^{kl}\big[a_i\partial_k\psi(\partial_j
a_l\!+\!f_{jl})\!-\!\partial_k\partial_i\psi a_j
a_l\big]_{\star_3}\big)\\&+\bar\psi\mathcal{O}(a^3)\psi
\end{split}
\end{equation}
One noticeable feature of $L_{pf}$ is that it contains vertices
identical to NCQED (without Seiberg-Witten map) in leading order
instead of ordinary QED as the $\theta$-expanded approach does. This
observation holds also for $L_{pp}$. One thus knows that the
one-loop two point function will contain UV/IR mixing terms coming
from those integrals in the same way as NCQED. The question is only
whether there will be new corrections coming from terms that arise
solely due to Seiberg-Witten map or not. As we will see in the next
section, for the fermion loop, the leading order IR divergent result
is fully identical for NCQED with and without Seiberg-Witten map.

\section{One-loop computation}
\begin{figure}[h]
\centering
\includegraphics[width=0.50\textwidth]{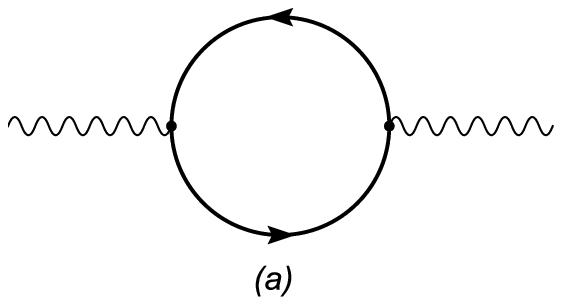}
\centering
\includegraphics[width=0.50\textwidth]{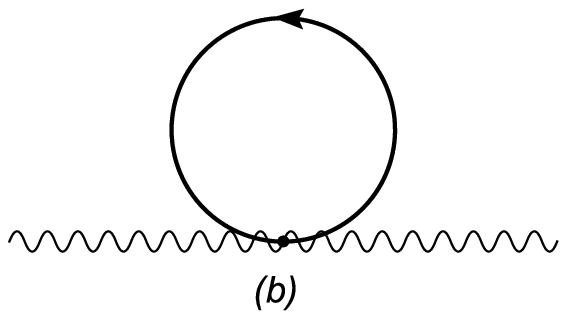}
\caption{Fermion loop corrections to the photon self
energy}\label{fermion:loop}
\end{figure}

The free part of the action $(\ref{ncqed:sw})$ is completely
identical to ordinary commutative QED, hence the quantization is
straightforward. Vertices coming from the fermion-photon interaction
lagrangian $(\ref{pf:int})$ are listed in the appendix. The fermion
loop contribution to the one loop photon two point function contains
two diagrams: the normal vacuum polarization graph as shown in as
shown in figure \ref{fermion:loop}(a) and a new fermion tadpole
graph in figure \ref{fermion:loop}(b).

Diagram (a) leads to following integral:
\begin{equation}\label{3-3-1}
\begin{split}
i\Pi^{ij}_{3-3}&=-4i\!\!\int
\frac{d^4k}{i(2\pi)^4}\frac{1}{(k+\frac{p}{2})^2(k-\frac{p}{2})^2}\sin^2\frac{p\wedge
k}{2}\tr\bigg\{[\gamma^i(\fmslash k+\frac{\fmslash
p}{2})\gamma^j(\fmslash k-\frac{\fmslash
p}{2})]\\&\!\!+\frac{1}{p\wedge k}[(\tilde p^i\fmslash k-\tilde
k^i\fmslash p)(\fmslash k+\frac{\fmslash p}{2})\gamma^j(\fmslash
k-\frac{\fmslash p}{2})+\gamma^i(\fmslash k+\frac{\fmslash
p}{2})(\tilde p^j\fmslash k-\tilde k^j\fmslash p)(\fmslash
k\!-\!\frac{\fmslash p}{2})]\\&\!\!+\frac{1}{(p\wedge k)^2}[(\tilde
p^i\fmslash k-\tilde k^i\fmslash p)(\fmslash k+\frac{\fmslash
p}{2})(\tilde p^j\fmslash k-\tilde k^j\fmslash p)(\fmslash
k-\frac{\fmslash p}{2})]\bigg\}
\end{split}
\end{equation}
where $\tilde p^i=\theta^{ij}p_j$.

For diagram (b) we get a surprising result: Its contribution can be
shown to vanish:
\begin{equation}
\begin{split}
i\Pi_4^{ij}&=i\int
\frac{d^4k}{i(2\pi)^4}\frac{1}{k^2}\tr\bigg\{4\frac{\sin^2\frac{p\wedge
k}{2}}{p\wedge k}(\tilde k^{i_1}\fmslash k\gamma^{i_2}+\tilde
k^{i_2}\fmslash k\gamma^{i_1})-4\frac{\sin^2\frac{p\wedge
k}{2}}{(p\wedge k)^2}(\fmslash k\fmslash p+\fmslash k\fmslash k
)\tilde k^{i_1}\tilde k^{i_2}\\&+\!2\fmslash k\fmslash
k\bigg[(-\!2\tilde p^{i_1}\tilde k^{i_2}\!+\!p\wedge
k\theta^{i_1i_2})\!+\!\frac{\sin^2\frac{p\wedge\!k}{2}}{(p\!\wedge\!k)^2}2(\tilde
k\!-\!\tilde p)^{i_1}\tilde
k^{i_2}\!+\!\frac{\sin^2\frac{p\wedge\!k}{2}}{p\!\wedge\!k}\theta^{i_1i_2}\!+\!\frac{\sin^2\frac{p\wedge\!k}{2}}{(p\!\wedge\!k)^2}(2\tilde
k^{i_2}\tilde p^{i_1}\\&+\theta^{i_1i_2}k\wedge
p)-\frac{\sin^2\frac{p\wedge k}{2}}{(p\wedge k)^2}\tilde
k^{i_1}\tilde k^{i_2}+(2\tilde p^{i_1}\tilde k^{i_2}-p\wedge
k\theta^{i_1i_2})+\frac{\sin^2\frac{p\wedge k}{2}}{(p\wedge
k)^2}2(\tilde k+\tilde p)^{i_1}\tilde
k^{i_2}\\&-\!\frac{\sin^2\frac{p\wedge\!k}{2}}{p\!\wedge\!k}\theta^{i_1i_2}\!-\!\frac{\sin^2\frac{p\wedge\!k}{2}}{(p\!\wedge\!k)^2}(2\tilde
k^{i_2}\tilde p^{i_1}\!+\!\theta^{i_1i_2}k\!\wedge
\!p)\!-\!\frac{\sin^2\frac{p\wedge\!k}{2}}{(p\!\wedge\!k)^2}\tilde
k^{i_1}\tilde
k^{i_2}\bigg]\!-\!4\frac{\sin^2\frac{p\wedge\!k}{2}}{p\!\wedge\!k}(\tilde
k^{i_1}\fmslash k\gamma^{i_2}\\&+\!\tilde k^{i_2}\fmslash
k\gamma^{i_1}\!)\!+\!4\frac{\sin^2\!\frac{p\wedge\!k}{2}}{(p\!\wedge\!k)^2}(\fmslash
k\fmslash p\!-\!\fmslash k\fmslash k)\tilde k^{i_1}\tilde
k^{i_2}\!+\!2\fmslash k\fmslash k\bigg[(2\tilde p^{i_2}\tilde
k^{i_1}\!\!-\!p\!\wedge\!
k\theta^{i_2i_1}\!)\!+\!\frac{\sin^2\!\frac{p\wedge\!k}{2}}{(p\!\wedge
\!k)^2}2(\tilde k\!+\!\tilde p)^{i_2}\tilde
k^{i_1}\\&-\!\frac{\sin^2\!\frac{p\wedge\!k}{2}}{p\!\wedge\!
k}\theta^{i_2i_1}\!-\!\frac{\sin^2\frac{p\wedge\!k}{2}}{(p\!\wedge\!
k)^2}(2\tilde k^{i_1}\tilde p^{i_2}\!+\!\theta^{i_2i_1}k\wedge
p\!)\!-\!\frac{\sin^2\!\frac{p\wedge\!k}{2}}{(p\!\wedge\!k)^2}\tilde
k^{i_1}\tilde k^{i_2}\!-\!(2\tilde p^{i_1}\tilde k^{i_2}\!-\!p\wedge
k\theta^{i_1i_2}\!)\\&+\!\frac{\sin^2\!\frac{p\wedge\!k}{2}}{(p\!\wedge\!k)^2}2(\tilde
k\!-\!\tilde p)^{i_2}\tilde
k^{i_1}\!\!+\!\frac{\sin^2\!\frac{p\wedge\!k}{2}}{p\!\wedge
\!k}\theta^{i_2i_1}\!\!-\!\frac{\sin^2\!\frac{p\!\wedge\!k}{2}}{(p\!\wedge\!k)^2}(2\tilde
k^{i_2}\tilde p^{i_1}\!\!+\!\theta^{i_1i_2}k\wedge
p)\!+\!\frac{\sin^2\!\frac{p\wedge\!k}{2}}{(p\!\wedge\!k)^2}\tilde
k^{i_1}\tilde k^{i_2}\!\bigg]\bigg\}\\&=i\int
\frac{d^4k}{i(2\pi)^4}\frac{1}{k^2}tr\bigg\{-8\frac{\sin^2\frac{p\wedge
k}{2}}{(p\wedge k)^2}\fmslash k\fmslash k\tilde k^{i_1}\tilde
k^{i_2}+8\frac{\sin^2\frac{p\wedge k}{2}}{(p\wedge k)^2}\fmslash
k\fmslash k\tilde k^{i_1}\tilde k^{i_2}\bigg\}\\&=0
\end{split}
\end{equation}
Hence we only need to evaluate the integral (\ref{3-3-1}). We work
out the trace in (\ref{3-3-1}), then write the wedge product in its
explicit component form, to obtain:
\begin{equation}\label{3-3-2}
\begin{split}
i\Pi^{ij}&=-16i\int
\frac{d^4k}{i(2\pi)^4}\frac{1}{(k+\frac{p}{2})^2(k-\frac{p}{2})^2}\sin^2\frac{p_i\theta^{ij}k_j}{2}\bigg\{[2k^ik^j\!-\!k^2g^{ij}\!-\!\frac{1}{4}(2p^ip^j\!-\!p^2g^{ij})]\\&-\!\frac{1}{p_i\theta^{ij}k_j}[2(p\cdot
k)(\tilde k^ik^j\!+\!k^i\tilde
k^j)\!-\!(k^2\!+\!\frac{p^2}{4})(\tilde p^ik^j\!+\!k^i\tilde
p^j\!+\!\tilde k^ip^j\!+\!p^i\tilde k^j)\!+\!\frac{1}{2}(p\cdot
k)(\tilde p^ip^j\\&+p^i\tilde
p^j)]+\frac{1}{(p_i\theta^{ij}k_j)^2}[(k^4-\frac{(p\cdot
k)^2}{2}+\frac{p^2k^2}{4})\tilde p^i\tilde p^j-(p^2k^2-2(p\cdot
k)^2+\frac{p^4}{4})\tilde k^i\tilde
k^j\\&-(k^2-\frac{p^2}{4})(p\cdot k)(\tilde p^i\tilde k^j+\tilde
k^i\tilde p^j)]\bigg\}
\end{split}
\end{equation}
As expected, we have here in the first square bracket terms that are
identical to ordinary NCQED. In the next pair of square brackets are
the new contribution coming from the Seiberg-Witten map together
with the non-trivial IR-divergent coefficients
$1/(p_i\theta^{ij}k_j)^n$, where $n$ equals to one for the second
and two for the third term. The integral in   seems to be not very
different to its counterpart in normal NCQED. Previous results
\cite{Hayakawa:1999zf,Matusis:2000jf} suggest that one can rewrite
\begin{equation}
\sin^2\frac{p_i\theta^{ij}k_j}{2}=\frac{1}{2}(1-\cos(p_i\theta^{ij}k_j))
\end{equation}
to separate terms with and without nontrivial phase shift (planar
and non-planar). However, the IR-divergent term
$1/(p_i\theta^{ij}k_j)^n$ introduces unexpected difficulties to the
usual re\-normalization procedure. The term $1/(p_i\theta^{ij}k_j)$
cannot be removed by introducing a Schwinger parameter as it does
not have a fixed sign in $R^4$. Furthermore, the term
$1/(p_i\theta^{ij}k_j)^2$ leads to a complicated Gaussian integral
over $k_\mu$ whose convergence in $R^4$ depends on the explicit
choice of $p_\mu$ (instead of $p^2$). Here, we try to evaluate the
leading order non-planar part by the following trick: We introduce
an additional variable $\lambda$ in the sine functions in
$(\ref{3-3-2})$ to make it $\sin \lambda(p_i\theta^{ij}k_j)$, then
one can cancel the negative power of $(p_i\theta^{ij}k_j)$ by taking
an appropriate number of derivative over $\lambda$, resulting
integral is:
\begin{equation}\label{3-3-3}
\begin{split}
i\Pi^{''ij}(\lambda)&=\!-8i\!\!\int
\frac{d^4k}{i(2\pi)^4}\frac{\cos(\lambda
p_i\theta^{ij}k_j)}{(k+\frac{p}{2})^2(k-\frac{p}{2})^2}\bigg\{(p_i\theta^{ij}k_j)^2[2k^ik^j\!-\!k^2g^{ij}\!-\!\frac{1}{4}(2p^ip^j\!-\!p^2g^{ij})]\\&-(p_i\theta^{ij}k_j)[2(p\cdot
k)(\tilde k^ik^j+k^i\tilde k^j)-(k^2+\frac{p^2}{4})(\tilde
p^ik^j+k^i\tilde p^j+\tilde k^ip^j+p^i\tilde
k^j)\\&+\!\frac{1}{2}(p\cdot\!k)(\tilde p^ip^j\!+\!p^i\tilde
p^j)]\!+\![(k^4\!-\!\frac{(p\cdot\!k)^2}{2}\!+\!\frac{p^2k^2}{4})\tilde
p^i\tilde
p^j\!-\!(p^2k^2\!-\!2(p\cdot\!k)^2\!+\!\frac{p^4}{4})\tilde
k^i\tilde k^j\\&-(k^2-\frac{p^2}{4})(p\cdot k)(\tilde p^i\tilde
k^j+\tilde k^i\tilde p^j)]\bigg\}
\end{split}
\end{equation}
The computation now proceeds along the lines of the standard
dimensional regularization method. Taking the derivative with
respect to $\lambda$ the integral (3.5) appears to be more divergent
than (3.5), fortunately the effective UV regulator coming from the
cosine decays exponentially and therefore is still effective. Now
one integrates over $\lambda$ and evaluates the resulting function
at $\lambda=1$. The free integration constant can be fixed by
matching the result for the first square bracket to the direct
computation in NCQED. Finally we obtained the following
(surprisingly simple) results for the leading order IR divergent
term:
\begin{equation}
\Pi^{ij}_{non-planar}=-\frac{64}{16\pi^2}\frac{\tilde p^i\tilde
p^j}{\tilde p^4}
\end{equation}
which is identical to the corresponding result in normal NCQED.
\section{Conclusion}
By explicit computation we have shown that NCQED defined via
Seiberg-Witten map still exhibits UV/IR mixing in its photon
one-loop two-point function, when this theory is treated
non\-perturbative\-ly in $\theta$. The proof of principle that this
non\-perturbative computation can be done at all is perhaps the most
important result of this work. To find the full expression for the
UV/IR mixing term one needs to compute also the photon self
interaction loop corrections, which can be done by a procedure
practically identical to the computation of the fermion loop. By the
arguments given in section~2, we know that there exists in general
also UV/IR mixing terms in the photon loop correction. Hence it is
quite safe to say that UV/IR mixing still exists in noncommutative
quantum gauge theories constructed using Seiberg-Witten maps and one
still needs to worry about unusual large modifications to the very
low energy physics from arbitrarily small $\theta$ since the
$\theta\rightarrow0$ limit is discontinuous at the quantum level.

Besides UV/IR mixing, the $\theta$-exact approach gives rise to a
regularization problem, which requires some improvement in the
re\-normalization procedure. From this view point the
$\theta$-expansion method in \cite{Bichl:2001cq} seems to be more
convenient. Another possible candidate is the Hamiltonian approach
to the re\-normalization, which has successfully achieved finite
results for noncommutative scalar field theory in Minkowski
space-time\cite{Bahns:2004hf}, while a related
approach\cite{Zahn:2006mg} to NCQED based on the Yang-Feldman
equation encountered similar problems for the photon two point
function as we encountered here\footnote{It is worth also to mention
that in \cite{Zahn:2006mg} an inexplicit expansion of open-Wilson
lines is constructed up to arbitrary formal order of the gauge
field, while the author probably did not notice the connection
between the expansion of open-Wilson lines and Seiberg-Witten maps
and erroneously claims that the Seiberg-Witten map is only valid in
an $\theta$-expanded way.}.

\acknowledgments Helpful discussions with Robert C. Helling are
gratefully acknowledged.\nobreak
\bigskip
\appendix
\section{Feynman rules for photon-fermion interaction}

\begin{figure}[!ht]
\centering
\includegraphics[width=0.30\textwidth]{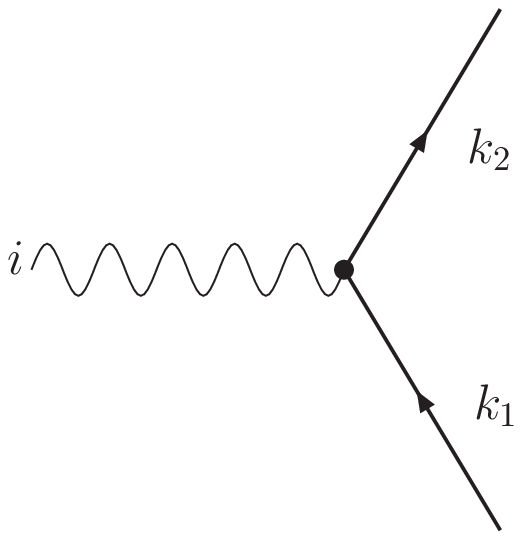}
\end{figure}
\begin{equation}
V^i_{pff}(k_1,k_2)=2\gamma^i\sin\frac{k_1\wedge k_2}{2}+2(\tilde
k_1^i\fmslash k_2-\tilde k_2^i\fmslash k_1)\frac{\sin\frac{k_1\wedge
k_2}{2}}{k_1\wedge k_2}
\end{equation}

\begin{figure}
\centering
\includegraphics[width=0.30\textwidth]{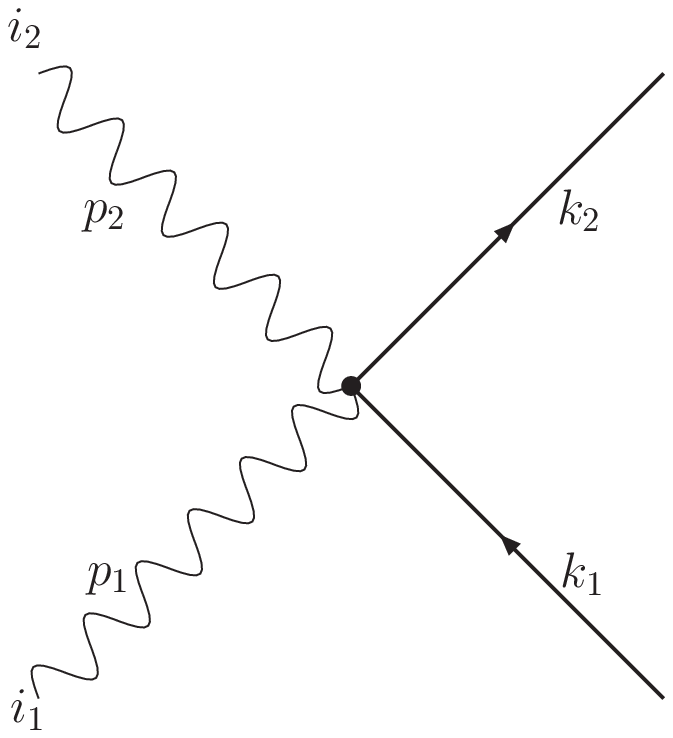}
\end{figure}
\begin{equation}
\begin{split}
V^{i_1i_2}_{ppff}(p_1,p_2,k_1,k_2)&=\bigg\{4i\frac{\sin\frac{p_1\wedge
k_1}{2}\sin\frac{p_2\wedge k_2}{2}}{p_1\wedge k_1}\tilde
k_1^{i_1}\gamma^{i_2}-4i\frac{\sin\frac{p_1\wedge
k_1}{2}\sin\frac{p_2\wedge k_2}{2}}{p_2\wedge k_2}\tilde
k_2^{i_2}\gamma^{i_1}\\&-2i\frac{\sin\frac{k_1\wedge
k_2}{2}\sin\frac{p_1\wedge p_2}{2}}{p_1\wedge
p_2}(2\gamma^{i_2}\tilde p_2^{i_1}-\fmslash
p_2\theta^{i_1i_2})-4i\frac{\sin\frac{p_1\wedge
k_1}{2}\sin\frac{p_2\wedge k_2}{2}}{p_1\wedge k_1p_2\wedge
k_2}(\fmslash p_2\\&+\fmslash k_2)\tilde k_1^{i_1}\tilde
k_2^{i_2}+2i\fmslash k_2\bigg[\frac{\sin\frac{k_1\wedge
k_2}{2}\sin\frac{p_1\wedge p_2}{2}}{p_1\wedge p_2k_1\wedge
k_2}(p_2\wedge k_1\theta^{i_1i_2}-2\tilde p_2^{i_1}\tilde
k_1^{i_2})\\&-\frac{\sin\frac{p_1\wedge k_2}{2}\sin\frac{p_2\wedge
k_1}{2}}{p_1\wedge k_2p_2\wedge k_1}2(\tilde p_2-\tilde
k_1)^{i_1}\tilde k_1^{i_2}+\frac{\sin\frac{p_1\wedge
k_2}{2}\sin\frac{p_2\wedge k_1}{2}}{p_1\wedge
k_2}\theta^{i_1i_2}\\&+\bigg(\frac{\sin\frac{p_2\wedge
k_1}{2}\sin\frac{p_1\wedge k_2}{2}}{p_2\wedge k_2p_1\wedge
k_2}+\frac{\sin\frac{p_1\wedge p_2}{2}\sin\frac{k_1\wedge
k_2}{2}}{p_2\wedge k_2k_1\wedge k_2}\bigg)(2\tilde k_1^{i_2}\tilde
p_2^{i_1}+\theta^{i_1i_2}k_1\wedge p_2\\&-\tilde k_1^{i_1}\tilde
k_1^{i_2})\bigg]+2i\fmslash k_1\bigg[\frac{\sin\frac{k_2\wedge
k_1}{2}\sin\frac{p_1\wedge p_2}{2}}{p_1\wedge p_2k_2\wedge
k_1}(2\tilde p_2^{i_1}\tilde k_2^{i_2}-p_2\wedge
k_2\theta^{i_1i_2})\\&+\frac{\sin\frac{p_1\wedge
k_1}{2}\sin\frac{p_2\wedge k_2}{2}}{p_1\wedge k_1p_2\wedge
k_2}2(\tilde p_2+\tilde k_2)^{i_1}\tilde
k_2^{i_2}-\frac{\sin\frac{p_1\wedge k_1}{2}\sin\frac{p_2\wedge
k_2}{2}}{p_1\wedge
k_1}\theta^{i_1i_2}\\&-\bigg(\frac{\sin\frac{p_2\wedge
k_2}{2}\sin\frac{p_1\wedge k_1}{2}}{p_2\wedge k_1p_1\wedge
k_1}+\frac{\sin\frac{p_2\wedge p_1}{2}\sin\frac{k_2\wedge
k_1}{2}}{p_2\wedge k_1k_2\wedge k_1}\bigg)(2\tilde k_2^{i_2}\tilde
p_2^{i_1}+\theta^{i_1i_2}k_2\wedge p_2\\&+\tilde k_2^{i_1}\tilde
k_2^{i_2})\bigg]+\{p_1\leftrightarrow p_2\,\mbox{and}\,
i_1\leftrightarrow i_2\}\bigg\}\delta(k_1-k_2-p_1-p_2)
\end{split}
\end{equation}
\bibliographystyle{JHEP}
\bibliography{uvirfermion}

\providecommand{\href}[2]{#2}\begingroup\raggedright\begin{thebibliography}{10}

\bibitem{Seiberg:1999vs}
N.~Seiberg and E.~Witten, {\it {String theory and noncommutative geometry}},
  {\em JHEP} {\bf 09} (1999) 032,
  [\href{http://xxx.lanl.gov/abs/hep-th/9908142}{{\tt hep-th/9908142}}].

\bibitem{Jurco:2001my}
B.~Jurco, P.~Schupp, and J.~Wess, {\it {Nonabelian noncommutative gauge theory
  via noncommutative extra dimensions}},  {\em Nucl. Phys.} {\bf B604} (2001)
  148--180, [\href{http://xxx.lanl.gov/abs/hep-th/0102129}{{\tt
  hep-th/0102129}}].

\bibitem{Jackiw:2001jb}
R.~Jackiw and S.~Y. Pi, {\it {Covariant coordinate transformations on
  noncommutative space}},  {\em Phys. Rev. Lett.} {\bf 88} (2002) 111603,
  [\href{http://xxx.lanl.gov/abs/hep-th/0111122}{{\tt hep-th/0111122}}].

\bibitem{Madore:2000en}
J.~Madore, S.~Schraml, P.~Schupp, and J.~Wess, {\it {Gauge theory on
  noncommutative spaces}},  {\em Eur. Phys. J.} {\bf C16} (2000) 161--167,
  [\href{http://xxx.lanl.gov/abs/hep-th/0001203}{{\tt hep-th/0001203}}].

\bibitem{Bichl:2001gu}
A.~A. Bichl, J.~M. Grimstrup, L.~Popp, M.~Schweda, and R.~Wulkenhaar, {\it
  {Deformed QED via Seiberg-Witten map}},
  \href{http://xxx.lanl.gov/abs/hep-th/0102103}{{\tt hep-th/0102103}}.

\bibitem{Jurco:2001rq}
B.~Jurco, L.~Moller, S.~Schraml, P.~Schupp, and J.~Wess, {\it {Construction of
  non-Abelian gauge theories on noncommutative spaces}},  {\em Eur. Phys. J.}
  {\bf C21} (2001) 383--388,
  [\href{http://xxx.lanl.gov/abs/hep-th/0104153}{{\tt hep-th/0104153}}].

\bibitem{Calmet:2001na}
X.~Calmet, B.~Jurco, P.~Schupp, J.~Wess, and M.~Wohlgenannt, {\it {The standard
  model on non-commutative space-time}},  {\em Eur. Phys. J.} {\bf C23} (2002)
  363--376, [\href{http://xxx.lanl.gov/abs/hep-ph/0111115}{{\tt
  hep-ph/0111115}}].

\bibitem{Melic:2005fm}
B.~Melic, K.~Passek-Kumericki, J.~Trampetic, P.~Schupp, and M.~Wohlgenannt,
  {\it {The standard model on non-commutative space-time: Electroweak currents
  and Higgs sector}},  {\em Eur. Phys. J.} {\bf C42} (2005) 483--497,
  [\href{http://xxx.lanl.gov/abs/hep-ph/0502249}{{\tt hep-ph/0502249}}].

\bibitem{Melic:2005am}
B.~Melic, K.~Passek-Kumericki, J.~Trampetic, P.~Schupp, and M.~Wohlgenannt,
  {\it {The standard model on non-commutative space-time: Strong interactions
  included}},  {\em Eur. Phys. J.} {\bf C42} (2005) 499--504,
  [\href{http://xxx.lanl.gov/abs/hep-ph/0503064}{{\tt hep-ph/0503064}}].

\bibitem{Buric:2007qx}
M.~Buric, D.~Latas, V.~Radovanovic, and J.~Trampetic, {\it {Nonzero Z $\to$
  gamma gamma decays in the renormalizable gauge sector of the noncommutative
  standard model}},  {\em Phys. Rev.} {\bf D75} (2007) 097701.

\bibitem{Alboteanu:2007zz}
A.~M. Alboteanu, {\em {The noncommutative standard model: Construction beyond
  leading order in Theta and collider phenomenology}}.
\newblock PhD thesis, {W\"urzburg}, 2007.

\bibitem{Bichl:2001cq}
A.~Bichl {\em et.~al.}, {\it {Renormalization of the noncommutative photon
  self-energy to all orders via Seiberg-Witten map}},  {\em JHEP} {\bf 06}
  (2001) 013, [\href{http://xxx.lanl.gov/abs/hep-th/0104097}{{\tt
  hep-th/0104097}}].

\bibitem{Buric:2002gm}
M.~Buric and V.~Radovanovic, {\it {The one-loop effective action for quantum
  electrodynamics on noncommutative space}},  {\em JHEP} {\bf 10} (2002) 074,
  [\href{http://xxx.lanl.gov/abs/hep-th/0208204}{{\tt hep-th/0208204}}].

\bibitem{Buric:2005xe}
M.~Buric, D.~Latas, and V.~Radovanovic, {\it {Renormalizability of
  noncommutative SU(N) gauge theory}},  {\em JHEP} {\bf 02} (2006) 046,
  [\href{http://xxx.lanl.gov/abs/hep-th/0510133}{{\tt hep-th/0510133}}].

\bibitem{Buric:2006wm}
M.~Buric, V.~Radovanovic, and J.~Trampetic, {\it {The one-loop renormalization
  of the gauge sector in the noncommutative standard model}},  {\em JHEP} {\bf
  03} (2007) 030, [\href{http://xxx.lanl.gov/abs/hep-th/0609073}{{\tt
  hep-th/0609073}}].

\bibitem{Minwalla:1999px}
S.~Minwalla, M.~Van~Raamsdonk, and N.~Seiberg, {\it {Noncommutative
  perturbative dynamics}},  {\em JHEP} {\bf 02} (2000) 020,
  [\href{http://xxx.lanl.gov/abs/hep-th/9912072}{{\tt hep-th/9912072}}].

\bibitem{Hayakawa:1999yt}
M.~Hayakawa, {\it {Perturbative analysis on infrared aspects of noncommutative
  QED on R**4}},  {\em Phys. Lett.} {\bf B478} (2000) 394--400,
  [\href{http://xxx.lanl.gov/abs/hep-th/9912094}{{\tt hep-th/9912094}}].

\bibitem{Hayakawa:1999zf}
M.~Hayakawa, {\it {Perturbative analysis on infrared and ultraviolet aspects of
  noncommutative QED on R**4}},
  \href{http://xxx.lanl.gov/abs/hep-th/9912167}{{\tt hep-th/9912167}}.

\bibitem{Matusis:2000jf}
A.~Matusis, L.~Susskind, and N.~Toumbas, {\it {The IR/UV connection in the
  non-commutative gauge theories}},  {\em JHEP} {\bf 12} (2000) 002,
  [\href{http://xxx.lanl.gov/abs/hep-th/0002075}{{\tt hep-th/0002075}}].

\bibitem{Mehen:2000vs}
T.~Mehen and M.~B. Wise, {\it {Generalized *-products, Wilson lines and the
  solution of the Seiberg-Witten equations}},  {\em JHEP} {\bf 12} (2000) 008,
  [\href{http://xxx.lanl.gov/abs/hep-th/0010204}{{\tt hep-th/0010204}}].

\bibitem{Bahns:2004hf}
D.~Bahns, {\it {The ultraviolet-finite Hamiltonian approach on the
  noncommutative Minkowski space}},  {\em Fortsch. Phys.} {\bf 52} (2004)
  458--463, [\href{http://xxx.lanl.gov/abs/hep-th/0401219}{{\tt
  hep-th/0401219}}].

\bibitem{Zahn:2006mg}
J.~W. Zahn, {\it {Dispersion relations in quantum electrodynamics on the
  noncommutative Minkowski space}},
  \href{http://xxx.lanl.gov/abs/0707.2149}{{\tt 0707.2149}}.

\end{thebibliography}\endgroup
\end{document}